\documentclass[twocolumn,superscriptaddress,showpacs,preprintnumbers,amsmath,amssymb]{revtex4}
\usepackage{graphicx}
\usepackage{dcolumn}
\usepackage{bm}
\begin{document}
\title{Geometric Phase in Eigenspace Evolution of Invariant and Adiabatic Action Operators}
\author{Jeffrey C. Y. Teo}
\affiliation{Department of Physics and Center of Theoretical and
Computational Physics, The University of Hong Kong, Pokfulam Road,
Hong Kong, China}
\author{Z. D. Wang}
\affiliation{Department of Physics and Center of Theoretical and
Computational Physics, The University of Hong Kong, Pokfulam Road,
Hong Kong, China} \affiliation{National Laboratory of Solid State
Microstructures, Nanjing University, Nanjing, China}

\date{\today}
\begin{abstract}
The theory of geometric phase is generalized to a cyclic evolution
of the {\it eigenspace} of an invariant operator  with $N$-fold
degeneracy.
 The corresponding geometric phase is interpreted as a
holonomy inherited from the universal connection of a Stiefel
$U(N)$-bundle over a Grassmann manifold. Most significantly, for
an arbitrary initial state, this geometric phase captures the
inherent geometric feature of the state evolution. Moreover, the
geometric phase  in the evolution of the eigenspace of an
adiabatic action operator is also addressed, which is elaborated
by a pullback $U(N)$-bundle.

\end{abstract}
\pacs{03.65.Vf, 03.65.Ca}

\maketitle

\newpage

Quantum geometric phases have been attracting significant research
interests, and have found many applications in various fields
including potential energy surfaces, the molecular Aharonov-Bohm
effect, Bloch bands in condense matters, and quantum Hall effects
{\it etc.} \cite{BMKN,SW}. A pioneering theory of
 geometric phase was established by Berry for an adiabatic cyclic evolution of a
non-degenerate energy eigenstate \cite{Berry1}, and its holonomy
interpretation was given by Simon \cite{Simon}. Latter,  the
geometric phase was respectively generalized to the cases  in the
adiabatic evolution of degenerate energy eigenstates by Wilczek
and Zee \cite{WZ}, and in the  evolution of a cyclic quantum state
by Aharonov and Anandan (AA) \cite{AA}; the potential applications
were addressed even in quantum computation \cite{Paolo}. Very
recently, an important attempt was made to introduce
phenomenologically a geometric phase accumulated in the evolution
that preserves a classical action adiabatically \cite{LWN3}. Thus
it is fundamental and significant to generalize the geometric
phase in the evolution of eigenspaces of an adiabatic action
operator, and particularly in the evolution of degenerate
eigenspaces of an invariant operator in
 a system with a periodic
time-dependent Hamiltonian, where even if the initial state is an
eigenstate of the invariant operator this state may not evolve
cyclically and thus the existing AA-phase theory is not
applicable. In this paper,
we establish a rigorous theory for the geometric phase in the
cyclic evolution of degenerate eigenspaces of an invariant
operator, and provide a holonomy interpretation inherited from the
universal connection of a Stiefel $U(N)$-bundle over a Grassmann
manifold. Most importantly, for an arbitrary initial state, this
geometric phase captures the inherent geometric feature of the
state evolution. Moreover, we also describe the geometric phase in
a cyclic evolution of the eigenspace of an adiabatic action
operator.

{\it Geometric Berry and AA Phases} The Hamiltonian $\hat H(R)$ of
a quantum system considered by Berry\cite{Berry1} and
Simon\cite{Simon} is a linear Hermitian operator on a complex
Hilbert space \textbf{H}, and depends smoothly on an external
parameter $R$ in a parameter space $M$, which is a smooth
manifold. If the system is non-degenerate, we have a complete set
of normalized eigenstates $\left| {n;R} \right\rangle $, with the
energy eigenvalue $E_n(R)$ being a smooth function defined
globally on $M$. On the other hand, there are various choices of
$\left| {n;R} \right\rangle $.  The dependence of an
$n^\textnormal{th}$-level eigenstate on the parameter $R$ is
referred as an $interpretation$. Clearly, one may define a new
$interpretation$ $\left| {n;R} \right\rangle '$ by a Gauge
transformation. We here are interested in single-valued
$interpretations$ that depend smoothly on a local domain in the
parameter space.
Consider the evolution of a physical $pure$ state $\left| {\psi (t)}
\right\rangle \left\langle {\psi (t)} \right|$,
where the state ket $\left| {\psi (t)} \right\rangle$ is governed by the
Schr\"{o}dinger's equation, with a time dependent parameter
$R(t)$. For a cyclic evolution $C$ in the parameter $R$, i.e,
$R(0)$=$R(T)$, the adiabatic approximation \cite{Berry1} with the
initial condition $\left| {\psi (0)} \right\rangle = \left|
{n;R(0)} \right\rangle$ leads to the solution $\left| {\psi (t)}
\right\rangle \approx k_n (t)\left| {n;R(t)} \right\rangle$ with
$k_n (T)=\exp \left( {iD_n (T)+i\gamma _n (C)} \right)\quad$,
where $D_n (T)=-\int_0^T {E_n (R(t))/\hbar dt}$ and $\gamma _n
(C)=i\int_C {\left\langle {n;R} \right|d \, /dR \left| {n;R}
\right\rangle dR}$ are  the dynamic phase and geometric Berry
phase respectively \cite{Berry1}.

From any arbitrary local $interpretation$, one can construct a
trivial $U(1)$-bundle over the domain of the
$interpretation$ in $M$, where the fibre of each $R$ is
$U(1)\otimes \left| {n;R} \right\rangle $. Suppose that two
$interpretations$ are related by the Gauge transformation
 $g_{\alpha \beta} $ with $\left| {n;R} \right\rangle _\beta =g_{\alpha \beta}
(R)\left| {n;R} \right\rangle _\alpha$, which gives a way to
identify points between trivial bundles defined by the
$interpretations$. By attaching these local bundles together, we
obtain a principal $U(1)$-bundle \cite{Kobayashi,Greub} $P$ over
$M$.
Equivalently, we can also construct a
complex line bundle $L$ over $M$ to describe the system, where the
fibre space of each $R$ is given by $\mathbb{C}\otimes \left|
{n;R} \right\rangle $. This line bundle is the bundle induced from
$P$ by the representation
 of the natural inclusion $i:U(1)\hookrightarrow
\mathbb{C}$ \cite{Greub}. The geometry of the bundle is described
by the Berry-Mead connection written in the form of a set of local
connection 1-forms:
\begin{equation}
\label{eq6}
\theta _\alpha = \left\langle {n;R} \right|d\left| {n;R}
\right\rangle _\alpha .
\end{equation}
This connection  gives a horizontal lifting of a curve in $M$. The
lifting is just the evolution of the state without the dynamic
phase. If the evolution is cyclic, the Berry phase is interpreted
as a holonomy in the principal $U(1)$-bundle (or complex line
bundle) \cite{Kobayashi}.

For the non-adiabatic but cyclic evolutions of physical $pure$ states,
i.e. $\left| {\psi (0)} \right\rangle \left\langle {\psi (0)}
\right|=\left| {\psi (T)} \right\rangle \left\langle {\psi (T)}
\right|$ (or $\left| {\psi (T)} \right\rangle =\exp \left( i
\gamma (T) \right) \left| {\psi (0)} \right\rangle $ with a real
$\gamma$ as the total phase shift), Aharonov and Anandan \cite{AA}
considered the space of $pure$ states as a complex
projective Hilbert space $P$(\textbf{H}).
Choosing a time-dependent normalized ket $\left| {\phi (t)}
\right\rangle $ such that $\left| {\psi (t)} \right\rangle
\left\langle {\psi (t)} \right|=\left| {\phi (t)} \right\rangle
\left\langle {\phi (t)} \right|$ and $\left| {\psi (0)}
\right\rangle = \left| {\phi (0)} \right\rangle=\left| {\phi (T)}
\right\rangle  $,
although $\left| {\phi (t)} \right\rangle $ may not be a solution
of the Schr\"{o}dinger's equation, it helps to give  a key result:
$\gamma (T)=D(T)+\gamma _{AA} (C) $, where $D(T)=- \int_0^T
{\left\langle {\phi (t)} \right|\hat {H}(t)\left| {\phi (t)}
\right\rangle }dt/\hbar $ and $\gamma _{AA} (C)=i\int_C
{\left\langle {\phi (t)} \right|d \, /dt\left| {\phi (t)}
\right\rangle dt}$ are the dynamic phase  and geometric AA phase,
respectively.

The geometric picture of AA phase is the AA bundle \textit{$\eta
$}, also known as the Hopf bundle \cite{Greub}, which is a
principal $U(1)$-bundle over the projective Hilbert space
$P$(\textbf{H}). This bundle is the universal
$U(1)$-bundle \cite{Steenrod}. An equivalent description is the
canonical line bundle \cite{Greub} \textit{$\xi $} over the
projective Hilbert space induced from the Hopf bundle by using the
representation of inclusion $i:U(1)\hookrightarrow \mathbb{C}$. The geometry
of this bundle is given by the local connection 1-form known as
the universal connection \cite{Bohm,Narasimhan} of the universal
bundle:
\begin{equation}
\label{eq13}
\omega _\alpha =\left\langle {\phi _\alpha } \right|d\left| {\phi _\alpha }
\right\rangle .
\end{equation}
Now, the relationship between the geometric pictures of Berry
phase and AA phase \cite{Bohm} becomes clear. Although the
$interpretation$ of energy eigenstate $\left| {n;R} \right\rangle
$ is local, the function $f:M\to P(\textbf{H})$ can be extended
globally on the parameter space $M$, mapping $R\mapsto \left|
{n;R} \right\rangle \left\langle {n;R} \right|$. Obviously, this
map is differentiable, and the principal $U(1)$-bundle $P$ (the
induced line bundle $L$) is the pullback bundle \cite{Greub} of
the Hopf bundle \textit{$\eta $} (the canonical line bundle
\textit{$\xi $}) by $f$, i.e. $P=f^< \eta $ ($L=f^< \xi $), where
$f^<$ denotes the covariant pullback functor of bundles induced
from $f$. The Berry-Mead connection is the pullback of the
universal connection $\theta =f^< \omega $ \cite{Greub}.

{\it Invariant Operators and Related Geometric Phases}
If a Hermitian operator \textit{\^{I}}(\textit{t})
depends smoothly on time and satisfies
\begin{equation}
\label{eq16}
\frac{\partial \hat
{I}}{\partial t}-\frac{i}{\hbar }[\hat {I},\hat {H}]\equiv 0,
\end{equation}
it is an invariant operator of the system ~\cite{Lewis}.
Intriguing properties of an invariant operator include: (i) all
eigenvalues are time-independent; (ii) if the state $\left| {\psi
_0 } \right\rangle$ is an eigenstate of
\textit{\^{I}}(\textit{t}$_{0})$,  its evolution $\left| {I (t)}
\right\rangle =U(t;t_0 )\left| {\psi _0 } \right\rangle $ is
always an eigenstate of \textit{\^{I}}(\textit{t}) with the same
eigenvalue, where \textit{U}(\textit{t};\textit{t}$_{0})$ is the
time-evolving operator; and (iii) transitions among eigenspaces
specified by different eigenvalues are impossible. We now first
consider the non-degenerate case of the invariant operator.
Whenever the following periodic condition is satisfied,
\begin{equation}
\label{eq18} \hat {I}(0)=\hat {I}(T), \hat {H}(0)=\hat {H}(T)
\end{equation}
with at least one of $\hat {I}$ and $\hat {H}$ being
time-dependent, the evolution of its eigenstate is found to be
cyclic. Thus we can deal with the geometric phase in the evolution
of an eigenspace of the invariant operator in the framework of the
AA phase theory. In the present case, each eigenspace of the
invariant operator is a ray in the Hilbert space and hence is a
point in the projective Hilbert space, so that
the evolution of eigenspaces corresponds exactly to the evolution
of physical $pure$ states.

As a simple but interesting example, let us consider an electron
subject to a time-independent magnetic field. The Hamiltonian
reads $\hat {H}=\mu_B B\sigma _z $ with eigenstates $\left|
\xi_\pm \right\rangle$ and eigen-energies $\pm \mu_B B$. Denoting
the initial state of the system as one of the two orthonormal
states: $\left| \psi_\pm (0) \right\rangle=\pm \cos \theta \left|
\xi_\pm \right\rangle +\sin \theta \left| \xi_\mp \right\rangle$,
we have $\left| \psi_\pm (t)\right\rangle=e^{\pm i\omega_s
t/2}\left| \phi_\pm \right\rangle$, where
\begin{equation}
\left| \phi_\pm \right\rangle=\pm \cos \theta \left| \xi_\pm \right\rangle +e^{\pm i\omega_s t}\sin \theta \left| \xi_\mp \right\rangle,
\end{equation}
and $\omega_s =2\mu_B B/\hbar $.  Then, an invariant operator is
found to be $\hat {I}(t)=\left| \phi_+ \right\rangle\left\langle
\phi_+\right|-\left| \phi_- \right\rangle\left\langle
\phi_-\right| =\sin 2\theta \cos \omega _s t\sigma _x +\sin
2\theta \sin \omega _s t\sigma _y +\cos 2\theta \sigma _z$, where
 $\left| \phi_\pm \right\rangle$ are its eigenstates with eigenvalues $\pm 1$
 and $\sigma_{\alpha}$'s are the Pauli matrices.
The connection 1-forms are $\omega _\pm (t)=\left\langle \phi_\pm
\right| d \, / dt\left| \phi_\pm \right\rangle dt=\pm i\omega _s
(1-\cos 2\theta)dt/2$, and the geometric phases for one cyclic
evolution  are $\gamma _\pm =i\int_0^T {\omega _\pm } =\pi (1\pm
\cos 2\theta)$.

As a more complicated example, we consider
an electron in a one-dimensional ring with radius \textit{a} and
subject to a crown-shaped electric field $\textbf{E}(\varphi
)=E\sin \chi \cos \varphi \hat {e}_x +\sin \chi \sin \varphi \hat
{e}_y +\cos \chi \hat {e}_z $ \cite{Wang2}. The Hamiltonian reads
\begin{equation}
\label{Example2}
\hat {H}=\hbar \Omega \left[-i\frac{\partial }{\partial \varphi }+\hat s_\varphi \right]^2,
\end{equation}
where $\hat s_\varphi =-\epsilon /2(\cos \chi \cos \varphi \sigma _x
+\cos \chi \sin \varphi \sigma _y -\sin \chi \sigma _z )$,
$\epsilon={\mu_B Ea}/\hbar$, and $\Omega=\hbar/2ma^2$. The
corresponding non-degenerate energy eigenstates are
\begin{equation}
\label{Example2a}
\begin{array}{l}
 \left| {\xi_{n+}} \right\rangle =e^{in\varphi}\left(
{{\begin{array}{*{20}c}
 {\cos \Theta } \hfill \\
 {e^{i\varphi}\sin \Theta } \hfill \\
\end{array} }} \right), \\
 \left| {\xi_{n-}} \right\rangle =e^{in\varphi}\left(
{{\begin{array}{*{20}c}
 {-\sin \Theta} \hfill \\
 {e^{i\varphi}\cos \Theta} \hfill \\
\end{array} }} \right), \\
 \end{array}
\end{equation}
where $\tan 2\Theta=\Delta/g$ with $\Delta=\epsilon \cos \chi$ and
$g=1-\epsilon \sin \chi$.
Letting $\left| \psi_{n\pm}(0)\right\rangle=\pm \cos \theta_n
\left| \xi_{n\pm} \right\rangle +\sin \theta_n \left| \xi_{n\mp}
\right\rangle$, since the Hamiltonian is time-independent, we can
use the same scenario to obtain the invariant operator
\begin{equation}
\label{Example3} \hat {I}(t)=\sum_n \left[(I_n+1)\left| {\phi _{n+}}
\right\rangle \left\langle {\phi _{n+}} \right|
 +(I_n-1)\left| {\phi _{n-}} \right\rangle
\left\langle {\phi _{n-}} \right| \right],
\end{equation}
with its eigenstates being
\begin{equation}
\label{Example11}
\left|
{\phi _{n\pm }} \right\rangle
 = {\pm \cos \theta_n \left| {\xi _{n\pm } }
\right\rangle +e^{\pm i\Omega _{ns} t}\sin \theta_n \left| {\xi
_{n\mp } } \right\rangle },
\end{equation}
where $\Omega _{ns}=\Omega (n+1/2)\sqrt
{\Delta^2+g^2}$ is the angular frequency for spin precession, and
$I_n$ can be chosen to be an arbitrary real number such that
the eigenvalues $I_n\pm 1$ are non-degenerate. The
connection 1-forms are $\omega _{n\pm } (t)=\pm i\Omega _{ns}
(1-\cos 2\theta_n)dt/2$, and the geometric phases for one cyclic
evolution of the eigenstates  are $\gamma _{n\pm } =\pi (1\pm \cos
2\theta_n)$.

At this stage, it is more significant and highly non-trivial to
tackle a general degenerate case of  the invariant operator
because a state may not evolve cyclically even if it is initially
an eigenstate of $\hat I$ and Eq.(\ref{eq18}) is satisfied. In
this case, supposing that the $n$-th level of the invariant
operator has an $N$-fold degeneracy, the $n$-th eigenspace  has a
complex dimension larger than one. Hence, the evolution of this
eigenspace can no longer be viewed as the evolution of a physical
$pure$ state in the projective Hilbert space, because the
eigenspace is not a ray but a higher dimensional subspace in the
Hilbert space. Indeed,  when Eq.(\ref{eq18}) is satisfied, each
eigenspace does undergo a cyclic evolution, rather than the
eigenstate. Hereafter, we define this kind of cyclic evolution of
the eigenspace of the invariant operator as the non-Abelian cyclic
evolution, with Eq.(\ref{eq18}) being its condition. A proper
geometric object tackling an $N$-fold degeneracy is the complex
Grassmann manifold $Gr(K;N)$ \cite{Greub}, where  $Gr(K;N)$ is the
space of $N$-dimension subspaces in the Hilbert space with $K$ as
the complex dimension  of the Hilbert space \cite{Remark2}.
Based on this picture, the
evolution of the eigenspace can be treated as a curve in
$Gr(K;N)$, and the evolution is non-Abelian cyclic only if the
curve is a loop. The corresponding geometric phase is non-Abelian.
In fact, as addressed
in Ref.\cite{Bohm}, the non-Abelian geometric phase
is interpreted as a holonomy of the Stiefel $U(N)$-bundle
$V_0(K;N)$\cite{Greub}, with $V_0(K;N)$ being the space of
orthogonal $N$-frames of the Hilbert space, over the Grassmann
manifold $Gr(K;N)$ with a universal connection
\cite{Bohm,Narasimhan}, which is written as a local connection
1-form $\omega _\alpha =\left( {{\omega _\alpha}_s^r }
\right)_{N\times N}$ with value inside the Lie algebra $u(N)$ of
skew-Hermitian matrices:
\begin{equation}
\label{eq20}
{\omega _\alpha}_s^r =\left\langle {{\phi _\alpha}_r } \right|d\left| {{\phi _\alpha}_s } \right\rangle ,
\end{equation}
where $\phi _\alpha $ is a local $section$ \cite{Kobayashi,Greub}
from the Grassmann manifold to the Stiefel bundle, mapping $V$ to
$\left( {{\phi _\alpha}_1 (V),\textit{{\ldots}},{\phi _\alpha}_N
(V)} \right)$ with  $\left( {{\phi _\alpha}_1
(V),\textit{{\ldots}},{\phi _\alpha}_N (V)} \right)$ as an
orthonormal $N$-frame that spans the space $V$.

An equivalent method
to address this issue is to use the canonical holomorphic vector
bundle $\Xi $ \cite{Greub} of rank $N$ over the Grassmann manifold
with the universal connection described by Eq.(\ref{eq20}). This
method seems more physical, since the fibre space over a point $V$
in the Grassmann manifold can be treated as the vector space $V$
itself. Upon the evolution of the invariant operator, a curve is
parameterized on the Grassmann manifold and the fibre space over
each point on the curve can be viewed as  the eigenspace of the
operator.

Let the non-Abelian cyclic evolution define a loop $t\mapsto V(t)$
in $Gr(K;N)$ that can be covered by the domain of
one local {\it section} $\phi _\alpha $. This {\it section} can
then be treated as a curve in the canonical vector bundle
parameterized by time in the following composition $\phi _\alpha
:t\mapsto V(t)\mapsto \phi _\alpha \left(V(t)\right)$, with
$V(0)=V(T)$ and $\phi_\alpha (0)=\phi_\alpha (T)$.
 The corresponding geometric phase is
\begin{equation}
\label{eq24}
\Gamma(C)=i\int_C {\omega _\alpha },
\end{equation}
where $\omega _\alpha $ is the 1-form
in Eq.(\ref{eq20}). Remarkably,
this geometric phase captures the inherent geometric feature of
the related evolution, even for an $arbitrary$ initial state. This
is because the state evolution can be decomposed into linear
combinations of the invariant operator eigenstates with
time-independent expansion-coefficients
 according to the property (ii) below Eq.(2) and thus the geometric phase
(matrix) can be represented by the direct sum $\Gamma=\oplus_n
\Gamma_n$ with $\Gamma_n$ as the geometric phase for the $n$-th
eigenspace. Therefore, the present invariant operator scenario
exhibits a superior advantage in exploring the geometric nature of
the non-Abelian cyclic evolution of the system with an arbitrary
initial state \cite{Note2}.

As an intriguing  illustration, let us still consider an electron
 in a ring, but subject to a
rotating electric field $\textbf{E}(\varphi ,t)=E\sin \chi \cos
(\varphi -\omega _o t)\hat {e}_x +\sin \chi \sin (\varphi -\omega
_o t)\hat {e}_y +\cos \chi \hat {e}_z $ with a non-zero
$\omega _o$. The Hamiltonian Eq.(\ref{Example2}) will be
time dependent by replacing $\hat {s}_\varphi $ with $\hat
{s}_{\varphi -\omega _o t}$. The two $n$-th level degenerate
eigenstates of the invariant operator $\hat {J}=-i\frac{\partial
}{\partial \varphi }+\frac{1}{2}\sigma _z$ with eigenvalue $n+1/2$
are
\begin{equation}
\label{Example2b}
\begin{array}{l}
 \left| {\xi'_{n+}} \right\rangle =e^{in\varphi}\left(
{{\begin{array}{*{20}c}
 {\cos \Theta } \hfill \\
 {e^{i(\varphi -\omega _o t)}\sin \Theta } \hfill \\
\end{array} }} \right), \\
 \left| {\xi'_{n-}} \right\rangle =e^{in\varphi}\left(
{{\begin{array}{*{20}c}
 {-\sin \Theta} \hfill \\
 {e^{i(\varphi -\omega _o t)}\cos \Theta} \hfill \\
\end{array} }} \right). \\
 \end{array}
\end{equation}
Using Eqs. (10)-(12), 
the non-Abelian geometric phase is found to be
the Hermitian matrix
\begin{equation}
\label{Example2d}
\Gamma_n=\left(
{{\begin{array}{*{20}c}
 {\pi (1-\cos 2\Theta)} \hfill & {\pi \sin 2\Theta} \hfill \\
 {\pi \sin 2\Theta} \hfill & {\pi (1+\cos 2\Theta)} \hfill \\
\end{array} }} \right).
\end{equation}
The matrix is independent of
$n$ and has both
diagonal and off-diagonal terms,  reflecting the spin precession
and flipping in the non-Abelian cyclic evolution.

{\it Adiabatic Action Operators and Geometric Phases} A linear
Hermitian operator \textit{\^{A}} is defined to be an action
operator if it commutes with the Hamiltonian. Here the action
operator \textit{\^{A}}(\textit{R}) and the Hamiltonian
\textit{\^{H}}(\textit{R}) are assumed to depend smoothly on an
external parameter \textit{R} in a parameter space \textit{M}, as
in the Berry phase case. When we consider the evolution of the
external parameter \textit{R}(\textit{t}) and if $\partial \hat
{A}(R(t))  \, / \partial t \approx 0$, the action operator $\hat
{A}$ is said to be approximately adiabatic in the evolution. Since
$[\hat {A}(R),\hat {H}(R)]=0$, $\hat {A}$   satisfies
Eq.(\ref{eq16}) approximately, namely it is an approximate
invariant operator.

The geometric phase of an eigenspace of the adiabatically evolving
action operator can be elaborated as follows.
By using the smooth function $f:M\to Gr(K;N)$, where
\textit{f}(\textit{R}) is the eigenspace of the action operator
\textit{\^{A}}(\textit{R}), we can define the pullback principal
bundle $P=f^< V_0(K;N)$ (or the vector bundle $\Pi=f^< \Xi $),
described by the local pullback connection
1-form $\theta_\alpha =f^< \omega_\alpha $.

To evaluate the geometric phase, we may choose an arbitrary smooth
local \textit{interpretation} $R\mapsto \left| {m,r;R}
\right\rangle _\alpha $ satisfying $f_> \left| {m,r;R}
\right\rangle _\alpha ={\phi _\alpha}_r \left( {f(R)} \right)$,
where $f_> :\Pi \to \Xi$ is the pushforward between bundles \cite{Greub},
\textit{m} denotes the eigenlevel of the action operator,
\textit{r} runs from one to \textit{N}. Hence, the
set of states $\left\{ {\left| {m,r;R} \right\rangle _\alpha }
\right\}_\alpha $ form an orthonormal basis of the
eigenspace \textit{f}(\textit{R}). Consequently, we can write the
connection for the bundles \textit{P} (or $\Pi$) explicitly as
$\theta _\alpha =f^< \omega _\alpha =\left( {{\theta _\alpha}_s^r
} \right)_{N\times N}$, where
\begin{equation} \label{eq32} {\theta _\alpha}_s^r = \left\langle
{m,r;R} \right|d\left| {m,s;R} \right\rangle _\alpha.
\end{equation}
The geometric phase is then
\begin{equation}
\label{eq33} \Gamma(C)=i\int_C {\theta _\alpha }.
\end{equation}

Interestingly, according to the Liouville's theorem in classical
mechanics \cite{Arnold}, the symplectic manifold representing a
periodic integrable system can be decomposed into leaves such that
each leaf is a submanifold diffeomorphic to a generalized torus,
and each leaf possesses a set of constants of motion:
\textbf{\textit{I}} \textit{= }\{\textit{I}$^{1}$,{\ldots},\,
\textit{I}$^{n}$\} with their conjugate variables being angles
\textbf{\textit{$\varphi $}}= \{\textit{$\varphi
$}$^{1}$,\textit{{\ldots}},\, \textit{$\varphi $}$^{n}\}$, where
\textit{$\varphi $}
  parameterize the \textit{n}-dimensional torus
$T^n$.
For the corresponding quantum system, let the wave function
$\left| {\psi (\textbf{\textit{I}})} \right\rangle =\psi
(\textbf{\textit{I}};\varphi )$ be an eigenfunction of an action
operator $\hat {A}$ with eigenvalues $\textbf{\textit{I}}$. The
inner product
is defined as $\left\langle {\psi (\textbf{\textit{I}})}
\mathrel{\left| {\vphantom {{\psi (\textbf{\textit{I}})} {\phi
(\textbf{\textit{I}})}}} \right. \kern-\nulldelimiterspace} {\phi
(\textbf{\textit{I}})} \right\rangle =\oint_{T^n} {\psi
(\textbf{\textit{I}};\varphi )^\dagger \phi
(\textbf{\textit{I}};\varphi )d\varphi }/(2\pi )^n$. We now
consider a non-degenerate case of the adiabatic action operator
depending on a slowly evolving external parameter \textit{R}.
For a cyclic evolution $C$ of the parameter, from Eq.(15), the
geometric phase is a real number \cite{Remark4}:
\begin{equation}
\label{eq36}
 \gamma (C)=\frac{1}{(2\pi )^n}\oint_{T^n} {\oint_C {\psi \left( {\textbf{\textit{I}}(R);\varphi }
 \right)^\dagger
\frac{\partial \psi }{\partial R}(\textbf{\textit{I}}(R);\varphi )dR} d\varphi }.
\end{equation}
This is just the geometric phase introduced phenomenologically in
Eq.(2) of Ref.\cite{LWN3}. We here present a rigorous derivation
with a precise geometric interpretation.

To  illustrate an application of Eq.(\ref{eq36}), we consider the
slowly evolving system described above Eq.(12) with a very small
non-zero $\omega_0$.
 The external parameter is
now the angle $\vartheta=\omega_0 t$ ($mod$ $2\pi$),
$\textbf{E}(\varphi ,t)=E\sin \chi \cos (\varphi -\vartheta)\hat
{e}_x +\sin \chi \sin (\varphi -\vartheta)\hat {e}_y +\cos \chi
\hat {e}_z $, and the parameter space of $\vartheta$ is the unit
circle $S^1$. The Hamiltonian is dependent on $\vartheta$ by
replacing $\hat {s}_\varphi$ by $\hat {s}_{\varphi-\vartheta}$ in
Eq.(\ref{Example2}). A natural choice of an action operator  is
$\hat {A}(\vartheta)=-i\frac{\partial }{\partial \varphi }+\hat
s_{\varphi -\vartheta}$.  The two eigenstates of the action
operator are
\begin{equation}
\label{Example3b}
\begin{array}{l}
 \left| {\xi'_{n+}(\vartheta)} \right\rangle =\xi'_{n+}(\vartheta ,\varphi)=e^{in\varphi}\left(
{{\begin{array}{*{20}c}
 {\cos \Theta } \hfill \\
 {e^{i(\varphi -\vartheta)}\sin \Theta } \hfill \\
\end{array} }} \right), \\
 \left| {\xi'_{n-}(\vartheta)} \right\rangle =\xi'_{n-}(\vartheta ,\varphi)=e^{in\varphi}\left(
{{\begin{array}{*{20}c}
 {-\sin \Theta} \hfill \\
 {e^{i(\varphi -\vartheta)}\cos \Theta} \hfill \\
\end{array} }} \right)
 \end{array}
\end{equation}
  with eigenvalues
$n+(1\pm \sqrt{\Delta^2+g^2})/2$. The connection 1-forms are
$\theta_{n\pm}(\vartheta)= \left\langle {\xi'_{n\pm}(\vartheta)}
\right|\frac{d}{d\vartheta}\left| {\xi'_{n\pm}(\vartheta)}
\right\rangle d\vartheta =-i(1\pm \cos 2\Theta)d\vartheta/2$.
Using Eq.(\ref{eq36}), the geometric phases of one cyclic
evolution of $\vartheta$ in the two eigenstates are found to be
\begin{equation}
\label{Example4}
\gamma_{n\pm} =\frac{1}{2\pi}\int_0^{2\pi} {\int_0^{2\pi} {{\xi'_{n\pm}}^\ast
\frac{\partial \xi'_{n\pm}}{\partial \vartheta}d\vartheta} d\varphi }
=\pi (1\mp \cos 2\Theta).
\end{equation}
Under this adiabatic approximation, the spin flipping is
neglected.

{\it Summary} We have developed a general theory of geometric
phase in the eigenspace evolution of invariant and adiabatic
action operators and elaborated its holonomy interpretation based
on the fiber bundle theory. Our theory has been applied to
non-trivial non-Abelian cases as well as to approximately
adiabatic action evolutions successfully. We anticipate that the
present theory will have more applications mainly because it is
applicable for any initial state of the system satisfying Eq.(4) .

We thank Q. Niu for useful discussions. The work was supported by
the RGC grant of Hong Kong, the URC fund of HKU, and the NSFC
grant(10429401).


\begin{thebibliography}{99}
\bibitem{Remark2} The dimension $K$ of the Hilbert space can be infinite.
\bibitem{Remark4}Eq.(\ref{eq36}) is proven (not shown here) to be also valid for
 a non-linear $U(1)$-invariant Hamiltonian(See, e.g., Ref.\cite{LWN3}).
\end{thebibliography}
\end{document}